\documentclass[aps,showpacs,twocolumn,amsmath,amssym,groupedaddress]{revtex4}
\usepackage{color,graphicx}
\usepackage{dcolumn}
\usepackage{bm}
\begin{document}
\title{Oscillatory dynamics of a charged microbubble under ultrasound}
\author{Thotreithem Hongray$^1$}
\author{B. Ashok$^2$}
\author{J. Balakrishnan$^{1,3}$}
\altaffiliation[Author to whom correspondence should be addressed. ]{Electronic mail: janaki05@gmail.com}
\affiliation{$^1$School of Physics, University of Hyderabad, Central Univ. PO, Gachi Bowli, Hyderabad 500046, India.}
\affiliation{$^2$International Institute of Information Technology, Bangalore (I.I.I.T.-B), 26/C, Electronics City, Hosur Road, Bangalore 560100, India.}
\affiliation{$^3$School of Natural Sciences \& Engineering, National Institute of Advanced Studies (N.I.A.S.), Indian Institute of Science Campus, Bangalore 560012, India.}
\begin{abstract} 
Nonlinear oscillations of a bubble carrying a constant charge and suspended in a fluid, 
undergoing periodic forcing due to incident ultrasound are studied. 
The system exhibits period-doubling route to chaos and the presence of 
charge has the effect of advancing these bifurcations.
The minimum magnitude of the charge $Q_{min}$ above which the bubble's radial oscillations 
can occur above a certain velocity $c_1$ is found to be related by a 
simple power law to the driving frequency $\omega$ of the acoustic wave. We find the existence of 
a critical frequency $\omega_H$ above which uncharged 
bubbles necessarily have to oscillate at velocities below $c_1$. We further find that this 
critical frequency crucially depends upon the amplitude $P_s$ of the driving acoustic pressure 
wave. The temperature of the gas within the bubble is calculated.  
A critical value $P_{tr}$ of $P_s$ equalling the upper transient threshold pressure 
demarcates two distinct regions of $\omega$ dependence of the maximal 
radial bubble velocity $v_{max}$ and maximal internal temperature  $T_{max}$. Above this pressure, 
$T_{max}$ and $v_{max}$ decrease with increasing $\omega$ while 
below $P_{tr}$, they increase with $\omega$. The dynamical effects of the charge and of 
the driving pressure and frequency of ultrasound on the bubble are discussed.\\
\end{abstract}

\maketitle

\section{Introduction}
The stability and oscillations of a gas bubble suspended in a liquid under the influence 
of an acoustic driving pressure field in the ultrasonic frequency range have been 
the subject of a large volume of scientific literature~\cite{rayleigh,plesset1,plesset2, neppiras,plessetProsperetti,brennenbook,suslick, brennerRMP, alty1,alty2, shiran1, shiran2, kellerMiksis,kellerKolodner}. 
Studies on the system have been made from different viewpoints coming from its diverse 
applications and occurrences. 
Ultrasound is routinely used in medical ultrasonography including echocardiography, lithotripsy, phacoemulsification, use in treatment of cancer and for dental cleansing.  
Other significant applications of ultrasonic forcing of fluids in which studies of bubble dynamics 
and cavitation become very important are in sonochemistry, sonoluminescence, ultrasonic cleaning 
of materials, waste water treatment and in focussed energy weapons. 
Cavitation events which involve violent collapse of micron-sized bubbles in the fluid can cause 
immense damage to the surfaces they are in contact with. Studies of cavitation events in pumps, 
turbines, surfaces exposed to hydrodynamic flow, etc., continue to be of immense interest in 
industries and in technological designs of devices.

Rayleigh's study of bubble cavitation was motivated by the need to understand and explain the 
damage to ships' propellers~\cite{rayleigh}. 
Under ultrasonic forcing, the behaviour of a bubble in a fluid depends heavily upon its ambient 
radius and the amplitude and frequency of the driving sound field.  
Thus the bubble can show regular oscillatory behaviour which can be periodic or it can show highly 
irregular oscillations which are chaotic and of unpredictable amplitude. 
For applications where damage caused on surfaces due to bubble cavitation can be disastrous, 
such as in medicine, it is desirous to operate the sonic device in a ``safe'' regime, and / or to 
be able to have control over the bubble's motion. 
Often in biological systems, it is known that bubbles in fluids can be electrostatically charged. 
Studies of the dynamics governing the oscillations, growth and collapse of charged bubbles are 
therefore of immense relevance because of their prevalence in diverse applications and situations. 
Experimental and theoretical work on the presence of charge on gas bubbles in fluids goes back to, 
for example, the work of McTaggart, Alty and Akulichev~\cite{alty1,alty2,mctaggart,akulichev}, and 
more recently the work of Shiran and Watmough and Atchley~\cite{shiran1,shiran2,atchley}. None of 
the work, though, has addressed the issue of dynamics of a charged bubble under ultrasonic forcing.\\
\indent
It is interesting to know what effect the presence of electric charge on the bubble would have and 
see if the motion of such a charged bubble forced by ultrasound would vary significantly from that 
of an electrically neutral bubble in a fluid. This especially becomes of practical significance when 
we are looking at cavitation phenomena in fluids in real-life, be it in the context of cavitation 
in mechanical systems or in the case of bubbles in fluids in living tissue in a medical context. 
Apart from the work in~\cite{zharov}, we are not aware of any other studies in 
the literature of the dynamics of acoustically forced charged bubbles suspended 
in a fluid. Their work however used the value $4/3$ for the polytropic constant 
which entailed cancellation of all the charged terms; thus their work does not 
really address the issue of charge which it sets out to do. 
The extremely nonlinear nature of the system, and the presence of a large number of parameters 
do not facilitate a straightforward analysis and it becomes essential to take the aid of 
numerical methods to get an understanding of the dynamics governing the observed behaviour. 
In this work we report some studies on the dynamics of a charged bubble in a liquid (which we 
take to be water) when ultrasound is incident on it. We assume that heat transfer across the 
bubble takes place adiabatically, and the gas is a monatomic ideal gas. We therefore take the 
polytropic constant $\Gamma = 5/3$.\\

In Section II we discuss briefly the nature of the radial dynamics of a charged bubble. 
Starting with a modified Rayleigh-Plesset equation, we obtain the time series of the bubble radius 
as also of its radial velocity and temperature. We also calculate the phase portrait of the bubble, 
under different pressure regimes.\\ In Section III we discuss the pressure thresholds that 
influence bubble dynamics; we introduce the expansion-contraction ratio $\zeta$ which we had 
introduced in~\cite{ashok}  that enables us to locate the presence of the Blake and upper transient 
threshold pressures easily when plotted as a function of the driving pressure amplitude $P_s$. 
The effects of driving frequency $\omega$ and charge on $\zeta$ are demonstrated in the present 
work.\\ The influence of $P_s$ and $\omega$ on the bubble dynamics are investigated in detail 
in Section IV. We obtain an expression for the minimum charge required on a bubble for radial 
oscillations to occur at some velocity $c_1$, as also the dependence on the forcing pressure amplitude 
$P_s$ of the maximum forcing frequency $\omega_H$ at which an uncharged bubble will oscillate with 
velocity $c_1$.\\
We then obtain, in Section V, the bifurcation diagrams for the system with driving frequency as 
the control parameter, and also the bifurcation diagram with charge as the control parameter. 
We observe that the presence of charge on the bubble advances period-doubling bifurcations with 
driving frequency as control parameter. Increasing $P_s$ causes the advancement of period doubling 
and halving bifurcations for charged as well as uncharged bubbles, and bands of chaotic behavior 
are observed at large $P_s$. \\The effect of charge and driving frequency on the maximal temperature 
are discussed in Section VI. We note that the pressure regime in which the bubble is being forced 
(whether $P_s$ is above or below the upper transient threshold pressure) determines the frequency 
dependence of the temperature, and we obtain rough limits on the maximum charge a bubble may carry 
depending on its ambient radius.\\
We conclude the paper with a summary of the results in Section VII. 
\section{Radial dynamics of the charged bubble}
In real-life situations, bubbles in fluids often have some electric charge sticking to them. 
This has been seen in the case of gas bubbles in various liquids as well as for cavitation events 
in water. In our work, we adapt the procedures for describing cavitation and forced bubble 
oscillations (that has a long and extensive literature), to include the presence of charge.\\
Description of ultrasonically forced bubble motion in a fluid has been made through the 
Rayleigh-Plesset equation~\cite{rayleigh,plesset1,plesset2} and its variants~\cite{plessetProsperetti, 
brennenbook,brennerRMP,kellerMiksis,kellerKolodner,parlitz,lofstedt, fengLeal,hilgenfeldt} modified 
to take into account compressibility of the fluid or various other factors. 
Proceeding as we did in our earlier work~\cite{ashok}, we further modify the form of the 
Rayleigh-Plesset equation for the evolution in time of the bubble radius $R(t)$ employed by
~\cite{parlitz} to include the presence of a constant charge $Q$ on the bubble as follows~\cite{zharov}: 
\begin{eqnarray}
&&\left(\left(1-\frac{\dot{R}}{c}\right)R
+\frac{4\eta}{c\rho}\right)\ddot{R}=\frac{1}{\rho}\left(P_0 -P_v+\frac{2\sigma}{R_0}
-\frac{Q^2}{8\pi\epsilon R_0^4}\right)\nonumber\\
&&\times\left(\frac{R_0}{R}\right)^{3\Gamma}\left(1+\frac{\dot{R}}{c}(1-3\Gamma)\right)
-\frac{\dot{R}^2}{2}\left(3-\frac{\dot{R}}{c}\right)\nonumber\\
&&+\frac{Q^2}{8\pi\rho\epsilon R^4}\left(1-\frac{3\dot{R}}{c}\right)-\frac{2\sigma}{\rho R}
-\frac{4\eta}{\rho}\left(\frac{\dot{R}}{R}\right)\nonumber\\
&&-\frac{1}{\rho}(P_0-P_v+P_s\textrm{sin}(\omega t))\left(1+\frac{\dot{R}}{c}\right)
-\frac{R}{\rho c}P_s\omega \textrm{cos}(\omega t)\label{cke}
\end{eqnarray} 

$R_0$ denotes the ambient equilibrium radius of the bubble, $P_0$ the static pressure, 
and $P_v = 2.34$kPa,  the vapour pressure of the gas. We denote by $P_s$ and $\omega = 2\pi\nu$ 
respectively ($\nu$ being the driving frequency), the amplitude and angular frequency of the 
ultrasound forcing field. We consider water to be the liquid surrounding the bubble, and having 
density $\rho = 998 kg/m^3$, viscosity $\eta = 10^{-3} Ns/m^2$, surface tension $\sigma = 0.0725 N/m$, 
and the velocity of sound in the liquid $c = 1500 m/s$, $P_0 = 101 kPa$, $\Gamma$ is the polytropic 
index and $\epsilon = 85 \epsilon_0$, where $\epsilon_0$ is the vacuum permittivity.\\ 

The modified Rayleigh-Plesset equation above can be simplified and rewritten in dimensionless 
form~\cite{ashok} as 
\begin{eqnarray}
&&\left(1-\frac{\dot{r}}{c_*}\right)r\ddot{r} + F\ddot{r}+\frac{\dot{r}^2}{2}\left(3
-\frac{\dot{r}}{c_*}\right) \nonumber \\ 
& =& H\left(1-P_{*v}+M\right)\left(\frac{1}{r}\right)^{3\Gamma}\left(1+\frac{\dot{r}}{c_*}(1-3\Gamma)\right)\nonumber \\
&+&\frac{C}{r^4}\left(1-\frac{3\dot{r}}{c_*}\right)-S\frac{1}{r}-Fc_*\left(\frac{\dot{r}}{r}\right) \nonumber \\
&-&H\left(1-P_{*v}+P_{*s}\sin(\tau)\right)\left(1+\frac{\dot{r}}{c_*}\right)-H\frac{rP_{*s}}{c_*}\cos(\tau) \nonumber \\
\end{eqnarray}
 
where $r= R/R_0$, $\tau = \omega t$, $P_{*v} = P_v/P_0$, $P_{*s} = P_s/P_0$ and the overdot here 
corresponds to differentiation with respect to $\tau$, and where the following dimensionless constants 
have been used: 
\begin{eqnarray*}
&&c_*=\frac{c}{R_0\omega}; \,\,\,\, F=\frac{4\eta}{\rho R_0 c}; \,\,\,\,  H=\frac{P_0}{R_0^2\omega^2\rho};\\
&&M=\frac{1}{P_0}\left(\frac{2\sigma}{R_0}-\frac{Q^2}{8\pi\epsilon R_0^4}\right)\\
&&C=\frac{Q^2}{8\pi\epsilon R_0^6{\omega}^2\rho};\,\,\,\,S=\frac{2\sigma}{\rho R_0^3\omega^2}
\end{eqnarray*}

We have employed the dimensionless form of the equations for obtaining their numerical solutions. 
In all the expressions that follow, and in the numerical results shown in graphs, we have rescaled 
the quantities by the appropriate factors and only displayed the dimensional form for a physical grasp 
of the magnitudes of the quantities involved.\\
The presence of charge $Q$ counters the effect of surface tension, reducing its effective value, 
and induces several interesting changes to the dynamics of bubble oscillations. 
In a previous work~\cite{ashok} we had obtained for the charged bubble, the Blake threshold and 
radius and also some results for the upper transient threshold for cavitation. In the following 
sections, we describe some interesting consequences of the presence of charge on a bubble.\\ 

As the bubble expands and contracts, the surface charge density decreases or increases respectively. 
The presence of charge lowers the surface tension and for sub-micron sized bubbles, dominates over it, 
influencing the minimum and maximum  values of the radius and the maximum velocities achieved by 
the bubble, and changing its point of collapse. A charged bubble achieves higher temperatures 
within it than an uncharged one, the collapse of the bubble being more violent in the charged case.\\ 

The above results indicate that since the bubble oscillations are more energetic for the charged 
bubble, the temperature attained by the gas within the bubble during its oscillations, would 
be higher as well. To confirm this, we calculate the temperature using equation (\ref{eqF})~\cite{lofstedt}.
\begin{equation}
T(t) = T(0)\left(\frac{R_0^3 - h^3}{R^3 - h^3}\right)^{\Gamma -1},
\label{eqF}
\end{equation}
where $h$ is the van der Waals hard core radius for the gas, $h=R_0/8.86$ for Argon 
~\cite{hilgenfeldt}. 
This equation is obtained under the assumption that there is no exchange of heat from the 
gas to its surroundings, that the system is essentially adiabatic.\\
\begin{figure}[]
\includegraphics[height=10.2cm,width=8.3cm,angle=0]{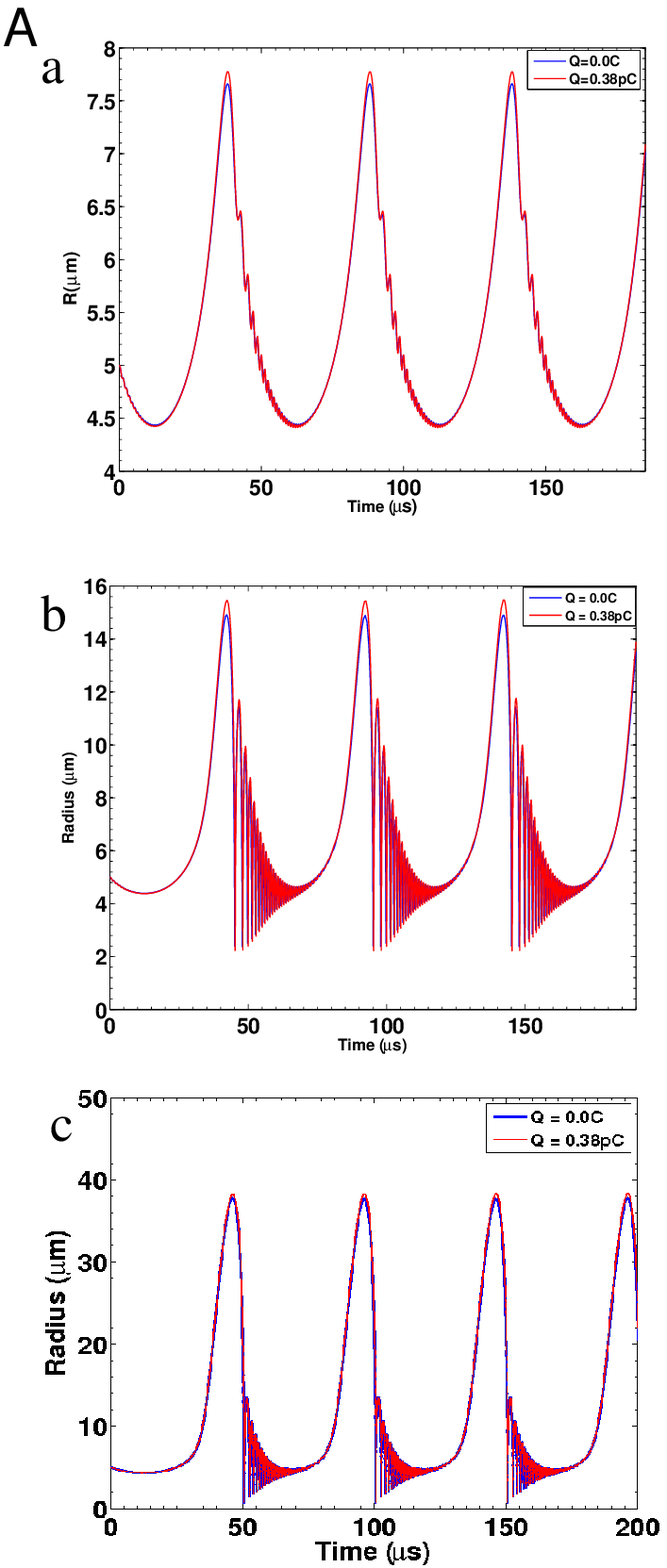}\\
\includegraphics[height=10.2cm,width=7.6cm,angle=0]{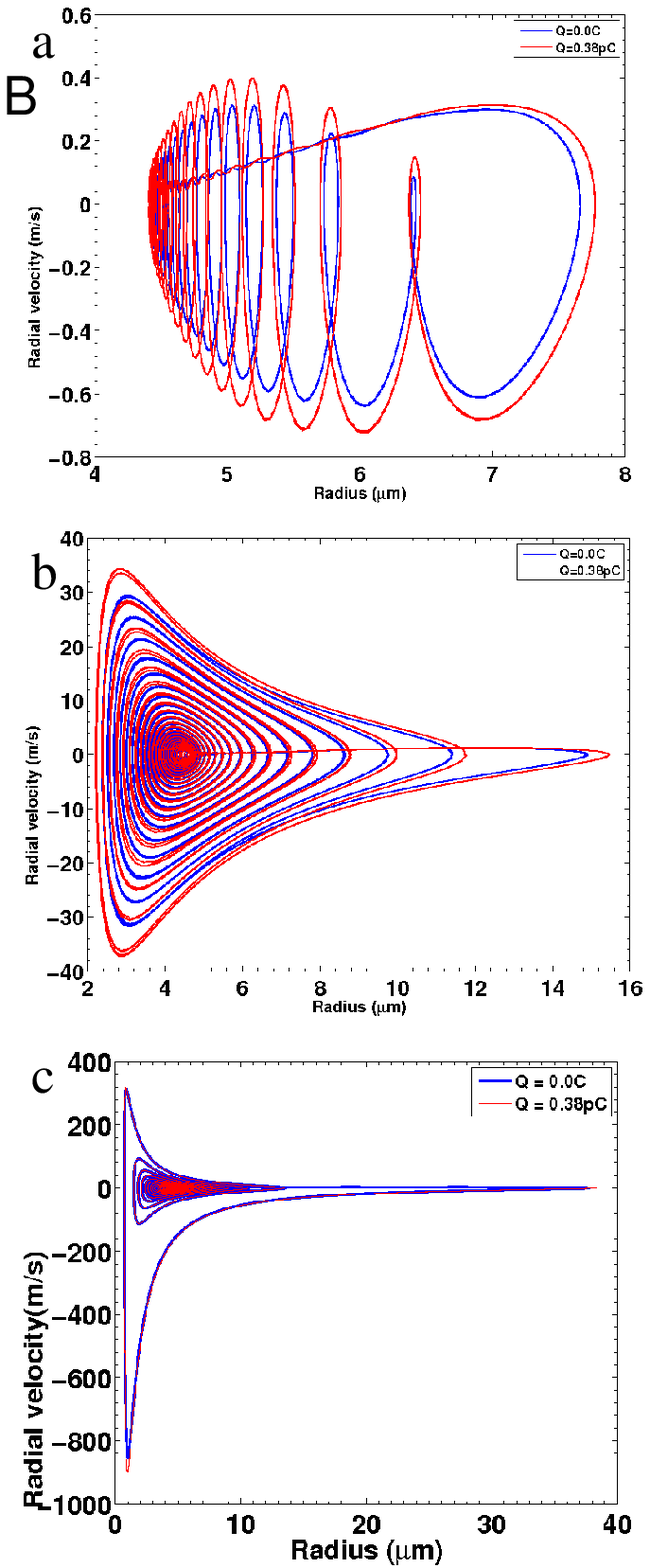}\\
\caption{A. Time series of $R$, and B. Phase portrait ($R$ vs $\dot R$), for $R_0 = 5 \mu$m 
and bearing charge $Q=0$(blue online) and $Q=0.38pC$ (red online),  
$P_s =$ (a) $1.0P_0$, (b) $1.12P_0$, (c) $1.25P_0$.  
Amplitudes of $R$, $\dot{R}$ are larger; the phase portrait spans a larger space for the charged 
bubble than for the uncharged case (color online).}
\label{fig1}
\end{figure} 
\begin{figure}[]
\includegraphics[height=9.5cm,width=8.5cm,angle=0]{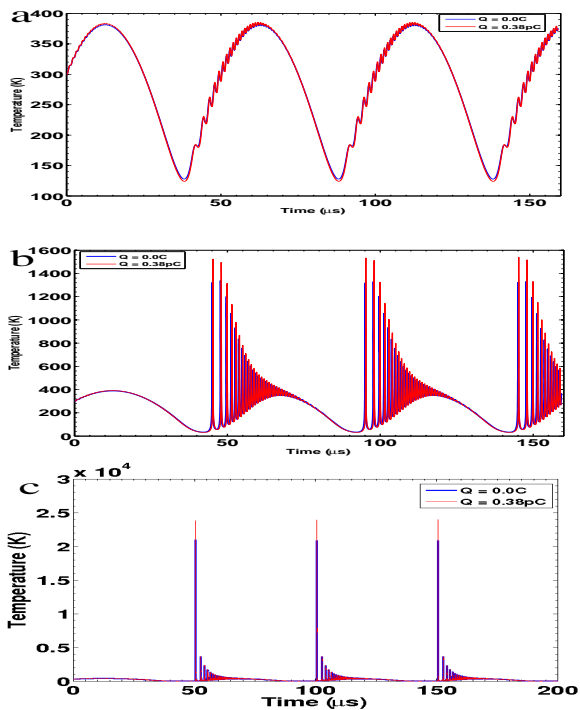}\\
\caption{Time series of temperature (in K) plotted as a function of time for 
$R_0=5\mu m$ at driving frequency $\nu = 20kHz$, and bearing charge $Q=0$(blue online) 
and $Q=0.38pC$ (red online)  for three values of $P_s$ : (a) $1.0P_0$, (b) $1.12P_0$, (c) $1.25P_0$. 
Temperatures $T$ are larger for the charged bubble than for the uncharged case (color online).}
\label{fig2}
\end{figure} 
 This assumption is not strictly true as in reality the equation of state of the gas enclosed within 
the bubble can be either adiabatic or isothermal, depending upon the rate of collapse of the bubble 
and whether or not the various relaxation time-scales permit thermal diffusion to occur to and from 
the bubble. We use the expression in order to get an idea of the magnitudes achievable by the 
temperature in the presence of charge. To better visualize the effect that the amplitude of the 
forcing pressure $P_s$ and charge $Q$ have on the bubble dynamics, we consider a bubble being 
driven at 20 kHz, i.e., the lower limit of the ultrasonic spectrum. 
Even in this lowest ultrasonic regime, the time series of bubble radius, radial velocity, and 
temperature all show an enhancement in values due to charge. Moreover, $P_s$ crucially determines 
the dynamics of the bubble as illustrated in Figures (\ref{fig1},\ref{fig2} A-B, a,b,c). 
We have considered three values of $P_s$, $P_s = 1.0 P_0$, $1.12 P_0$ and $1.25 P_0$. 
These pressures are, respectively, below the Blake threshold $P_{Blake}$, at the upper transient 
pressure threshold $P_{tr}$, and above $P_{tr}$. As can be seen, the pressure regime in which the 
bubble dynamics occurs, crucially determines the behavior. 
At $P_s = 1.0 P_0$, $T_{max} \approx 370 K$, the uncharged bubble temperature being marginally 
less than that for the charged bubble ($Q= 0.38$ pC); for $P_s = P_{tr} = 1.12 P_0$, $T_{max}$ 
goes up to about 1520 K for the charged bubble and about 1320 K for the uncharged case; and 
for $P_s = 1.25 P_0$, $T_{max}$ shoots up still further, to about 24,000 K for the charged 
(and approximately 21,000 K for the uncharged) bubble. These temperatures vary by orders of 
magnitude and spell out the importance of $P_s$ and $Q$. 

\section{Pressure thresholds}
The Blake threshold determines the pressure threshold beyond which an acoustically forced bubble 
undergoes drastic expansion. After the Blake threshold and preceding the onset of bubble collapse 
following a larger threshold known as the upper transient threshold, $P_{tr}$, the bubble is 
essentially in an unstable regime.\\ 
Depending upon whether the amplitude of the applied acoustic forcing pressure is greater or 
lesser than $P_{tr}$, the response of the bubble to the frequency of the applied pressure 
wave varies drastically.\\ 

At low amplitudes of the forcing pressure (i.e., $P_s < P_{tr}$), increasing driving frequency 
causes a proportional increase in the bubble's maximum radial velocity $v_{max}$. This happens 
upto some critical value of the frequency for that $P_s$ after which $v_{max}$ rises more steeply 
but accompanied with large oscillations.\\
At larger amplitudes of the forcing pressure, with $P_s$ approaching the 
value of the Blake and upper transient threshold pressures, the situation is 
different. $v_{max}$ first decreases with increasing driving frequency upto
a frequency $\omega_{hc}$, after which $v_{max}$ rises with frequency but 
with large oscillations. 
As could be expected from the above observations, a similar observation 
can be made regarding the maximum temperature $T_{max}$ of the gas inside the bubble.\\ 
A useful graphical illustration of the transient threshold pressures, i.e. of the Blake threshold 
($P_{Blake}$) and the upper transient threshold ($P_{tr}$) pressures, can be obtained by plotting 
$\zeta = (R_{max} - R_0)/(R_0 - R_{min})$ as a function of the amplitude of driving pressure. 
\begin{figure}[!h]
\includegraphics[height=5.2cm,width=8.5cm,angle=0]{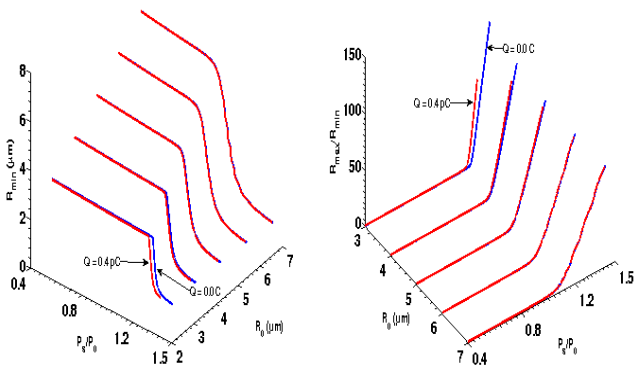}
\caption{Plots of $R_{min}$ (left) and $R_{max}/R_{min}$ (right) as functions of $P_s$ for 
different values of ambient radius (for $Q=0$ and $Q = 0.4$ pC) at 20kHz forcing frequency.
(color online).}
\label{fig3}
\end{figure}
This quantity $\zeta$, which we call the expansion-contraction ratio, handily shows the location 
of both the Blake and the upper transient thresholds. Both these thresholds cannot be identified 
easily at the same time from, for example, a plot of $R_{max}/R_{min}$ as a function of applied 
pressure amplitude. In Figure (\ref{fig3}), the points of inflection of the curves correspond 
to the Blake threshold pressures for the respective $R_0$ values. The effect of charge is clearly 
seen in reducing the threshold pressure as compared to the charged case. The upper transient 
threshold cannot be easily pinned down from this plot. While the Blake threshold is indicative 
of the threshold of the expansive growth of the bubble, the upper transient threshold demarcates 
where the violent collapse of the bubble occurs. A plot of $\zeta$  versus $P_s$ for different 
values of $R_0$ as shown in Figure (\ref{zetaplot}) shows a rise of the curve till it peaks 
(at $P_s = P_{Blake}$) followed by a trough or well (at $P_s = P_{tr}$) before rising up steeply 
for higher $P_s$ (this has been discussed in some detail in our earlier work~\cite{ashok}). 
At pressures between $P_{Blake}$ and $P_{tr}$ the bubble is in an unstable regime. This also 
explains the presence of large fluctuations or oscillations in the velocity vs. frequency plots 
at such intermediate pressures.
\begin{figure}[!h]
\includegraphics[height= 6cm,width=8cm,angle=0]{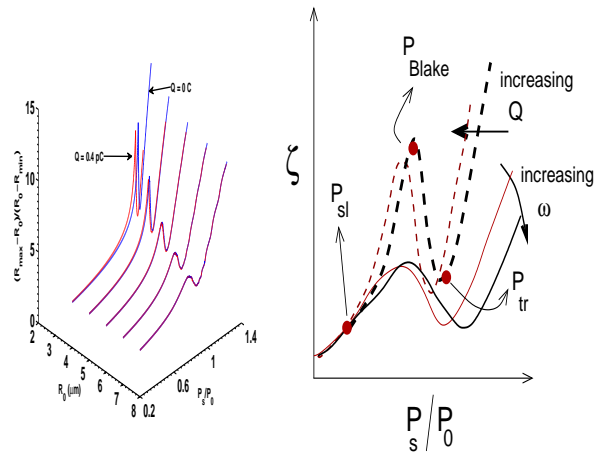}
\caption{Plot of $\zeta = (R_{max} - R_0)/(R_0 - R_{min})$ as a function of forcing pressure 
amplitude $P_s$ and $R_0$ at 20kHz. The schematic at right shows the locations of the Blake threshold 
pressure and the upper transient threshold pressure on the $\zeta - P_s$ curve. Increasing bubble 
charge shifts the curve down and to the left, while increasing driving frequency decreases the 
steepness of the curve and flattens it.(color online).}
\label{zetaplot}
\end{figure}
The presence of charge shifts the threshold pressures to lower values. With increasing ambient 
radius $R_0$, $\zeta$ loses its distinctive peak-valley appearance gradually.\\
The maximum radius attainable by the bubble gradually increases with charge for a given driving 
frequency~\cite{ashok}. 
This can be  understood from the fact that the presence of charge on the bubble decreases the 
effective surface tension. This causes the bubble to expand more easily in the negative pressure 
field. A casual reading might give rise to the observation that by the same argument, the 
minimum radius reached by the bubble would likewise follow a similar trend, with $R_{min}$ for 
a charged bubble having a larger value than that of a neutral bubble. However, this is not so. 
It should be borne in mind that $R_{min}$ is influenced by the maximal velocity the bubble is 
able to reach. The greater the velocity, the smaller the $R_{min}$ that it collapses to. 
Hence, perhaps counter-intuitively, charged bubbles undergoing forced oscillations, will achieve 
smaller values of $R_{min}$ than electrically neutral bubbles.\\
Thus presence of charge leading to greater bubble expansion, in turn results in the bubble collapse 
being much more rapid and violent, shrinking the bubble volume more than in the case of the uncharged 
bubble. This can be seen in Figure (\ref{fig3}) (left), where the minimum radius, $R_{min}$, reached 
by the bubble at the moment of collapse is plotted as a function of the driving pressure $P_s$ 
and $R_0$ for the charged and uncharged bubble. As was shown in greater detail in our earlier 
work~\cite{ashok}, $R_{min}$ reduces with increasing $Q$. 

\section{Influence of amplitude and frequency of driving pressure field} 
\begin{figure}[!h]
\includegraphics[height=13.5cm,width=8cm,angle=0]{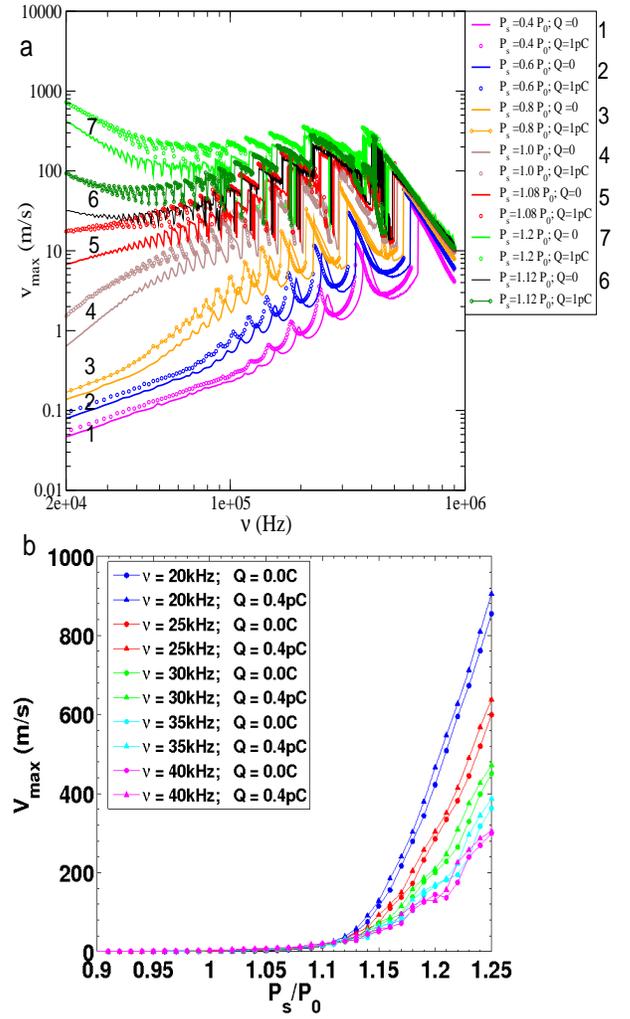}
\caption{Maximum velocity vs (a) driving frequency, plotted at 7 different $P_s$ values and for 
zero  \& non-zero  values of charge; the $P_s = 1.12 P_0$ curves (number 6, darkest ($Q=0$ black online) 
\& just above that ($Q=1$pC, dark green online)) correspond to the upper transient threshold pressure 
$P_{tr}$ (for $R_0 = 5 \mu m$) and clearly demarcate distinct behaviour for $P_s > P_{tr}$ and 
$P_s < P_{tr}$. (b) $v_{max}$ vs. $P_s$ for different frequencies for $R_0 = 5 \mu m$, at 
5 different driving frequencies (20 to 40 kHz, increasing from left to right), for zero (−$\bullet$−) 
and non-zero ($Q= 0.4$pC ,triangle) charge. (color online).}
\label{Rdotmaxplots}
\end{figure}

The maximum radial velocity of the bubble attained during its collapse or contracting phase 
depends also on the driving frequency, the charge present on the bubble, as well as the amplitude 
of the driving pressure wave, as also on the initial radius $R_0$ of the bubble in its quiescent 
state. Figure \ref{Rdotmaxplots} are plots of the maximal radial velocity as functions of the 
driving frequency (a) and pressure $P_s$ (b). There are several interesting features evident from 
the figures.\\  
The behaviour of $v_{max}$ above $P_s=P_{tr}$ is different from that below it. The plots shown 
are for a bubble of $R_0=5\mu m$ for which $P_{tr}=1.12P_0$. Fig.(\ref{Rdotmaxplots}a) shows that 
at $P_s<P_{tr}$, $v_{max}$ increases as a function of driving frequency $\nu$ while for  
$P_s>P_{tr}$ Fig.(\ref{Rdotmaxplots}b), it decreases. Increasing the driving frequency induces 
instability by producing large amplitude oscillations. 

\begin{figure}[!h]
\includegraphics[height=9cm,width=9cm,angle=0]{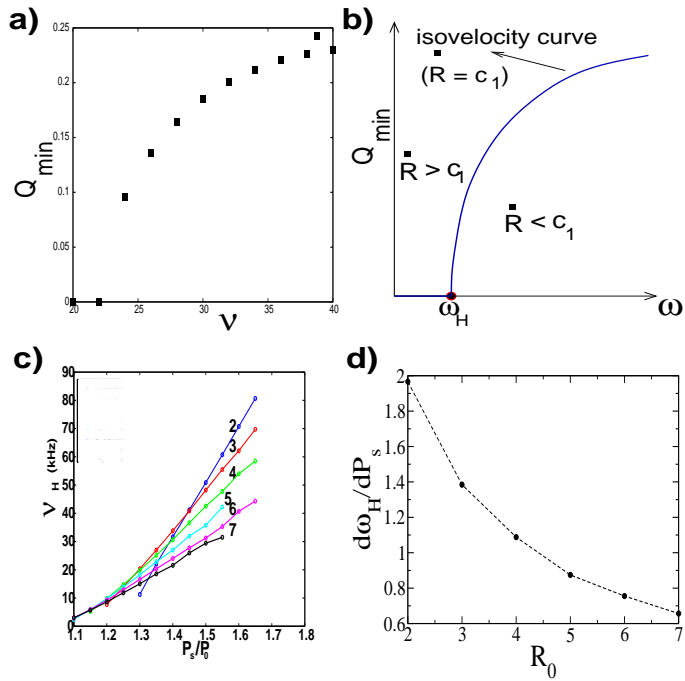}\\
\caption{$Q_{min}$ versus frequency $\omega$ iso-velocity plots ($v \approx 1500$ m/s) for 
a) $R_0 = 5 \mu m$, for $P_s = 1.35 P_0$  ($Q_{min}$ is in pC, $\omega$ in kHz); 
b) schematic illustrating the nature of the iso-velocity plot, 
$Q_{min} \sim (\omega - \omega_H)^{0.25}$; c) plot showing the dependence of $\omega_H$ on $P_s$ for 
different values of $R_0$ (curves 2,3,4,5,6,\& 7 correspond to corresponding $R_0$ values in $\mu$m);  
d) plot illustrating variation of ($d\omega_H/dP_s$) with $R_0$. (color online).}
\label{fig5}
\end{figure}

For a given magnitude of pressure amplitude $P_s$, the magnitude of charge present influences the 
dynamical regime of the bubble. If the driving angular frequency of the applied pressure wave 
is $\omega$ at a certain pressure amplitude, for bubble oscillations to occur with some maximal 
radial velocity $\dot{R} = c_1$, the charge present on the bubble should have some minimum magnitude 
$Q = Q_{min}(\omega)$. At low frequencies, even an uncharged bubble might oscillate at that velocity; 
however at higher frequencies, if charge $Q < Q_{min}$, the radial bubble velocity would be smaller 
than $c_1$. This is because as frequency increases, the bubble does not get sufficient time to 
complete its expansion, so that its subsequent collapse occurs with smaller radial velocity than if 
expansion to a greater size had been done. The presence of charge reduces the surface tension and 
encourages expansion to larger radial dimension and the consecutive, more violent collapse to a 
smaller radius. \\ 
That the change in $Q_{min}(\omega)$ with $\omega$ show a bifurcation in the parameter space is 
clear from Figure~\ref{fig5}. This transition from zero to non-zero $Q_{min}$ occurs at an angular 
frequency $\omega_H$. For $P_s= 1.35 P_0$ and $R_0 = 5 \mu$m, $\omega_H = 23 kHz$. 
The magnitude of $Q_{min}$ varies with driving frequency as
\begin{equation}
Q_{min} \approx a(\omega - \omega_H)^b,
\label{eq3}
\end{equation}
where the prefactor $a$ has appropriate dimensions and depends on the value of the initial ambient 
bubble radius $R_0$, and $b \approx 0.25$.\\ 
We could attempt to give a simple explanation for the frequency dependence of $Q_{min}$. 
We could argue that for a given value of constant maximal radial velocity $c_1$, 
the kinetic energy of the bubble would scale as the electrostatic contribution 
$\sim Q^2/R_{min}$, so that $Q^2 \sim a_2 c_1 R_{min}$, $a_2$ being a prefactor with appropriate 
dimensions. Since for high applied pressures (that is, for values of $P_s$ above the Blake threshold 
pressure or of the order of or above the upper transient threshold pressure) we know that the minimum 
bubble radius scales as the two-fifths power of the driving frequency $R_{min} \propto \omega^{2/5}$, 
it would follow that $Q^2 \propto c_1 \omega^{2/5}$ so that $Q \propto \omega^{0.2} $, which is close 
but not equal to the observed exponent of 0.25. Hence, this argument, while it serves to give a lower 
bound for $Q_{min}$, is insufficient.\\ 
Proceeding more systematically therefore, we start by making a linearization of the Rayleigh-Plesset 
equation. Proceeding along the lines of~\cite{plessetProsperetti}, the driving sound pressure is 
introduced through a small perturbation $\alpha$, so that the total external field $P_{ext}$ can 
be written as:
\begin{equation}
P_{ext} = P_0(1-\alpha \cos\omega t)
\end{equation}  
The bubble oscillations $R(t)$ about the equilibrium radius $R_0$ can then be expressed as 
\begin{equation}
R = R_0(1 + x(t))
\end{equation}
where $x(t)$ is a small quantity of order $\alpha$. Substituting this equation in the 
Rayleigh-Plesset equation (1) and linearizing it, we get 
\begin{equation}
\ddot x + \beta \dot x + \omega_0^2 x = - F_{ext}
\label{eq6}
\end{equation}
where the damping coefficient $\beta$, natural frequency of oscillation $\omega_0$ of the bubble 
and $F_{ext}$ are given by
\begin{eqnarray}
\beta &=& \frac{1}{\rho c R_0\left(1 + \frac{4\eta}{c\rho R_0}\right)}\left(4\phi/R_0^5 
+ \frac{3Q^2}{8\pi\epsilon R_0^4} + \frac{4\eta c}{R_0}\right) \nonumber \\
\omega_0^2 &=& \frac{1}{\rho R_0^2\left(1 + \frac{4\eta}{c\rho R_0}\right)}\left(5\phi/R_0^5 
-\frac{2\sigma}{R_0} + \frac{4Q^2}{8\pi\epsilon R_0^4}\right)
\end{eqnarray} 
Here only terms linear in $x$ and its derivatives have been retained and $\phi / R_0^5$ is 
the equilibrium gas pressure in the bubble defined by 
\begin{equation}
\phi / R_0^5 = ( P_0 - P_v + \frac{2\sigma}{R_0} - \frac{Q^2}{8\pi\epsilon R_0^4}) .
\end{equation} 
The particular situation of looking for conditions where the radial velocity is constant is 
thus implicitly satisfied. In eqn.(\ref{eq6}), $F_{ext}$ is given by
\begin{eqnarray}
F_{ext} &=& -P_s \frac{\left( 1 + \frac{R_0^2\omega^2}{c^2} \right)^{\frac{1}{2}} }
{\left( R_0^2 + \frac{4\eta R_o}{c\rho} \right)} \nonumber\\
&~&\times \cos \left( \omega t - \arctan \left(\frac{c}{R_0\omega}\right) \right) \nonumber\\
&\approx& -P_s \frac{\left( 1 + \frac{R_0^2\omega^2}{c^2} \right)^{\frac{1}{2}} }
{\left( R_0^2 + \frac{4\eta R_o}{c\rho} \right)}\cos \left(\omega t -\frac{\pi}{2}+
\frac{R_0\omega}{c}\right) 
\label{eq9}
\end{eqnarray}
where in arriving at the last line of eqn.(\ref{eq9}), use has been made of the fact 
that $\frac{R_0\omega}{c} \ll 1$. Scaling the time as $\hat t = \omega_0 t $ for convenience, 
eqn.(7) can be solved exactly. Dropping the hat ($\hat{}$) over $t$ for convenience of notation 
in all of the following, the steady state part of the solution is found to be 
\begin{eqnarray} 
x &=&-\frac{P_s\left( 1 + \frac{R_0^2\omega^2}{c^2} \right)^{\frac{1}{2}} }
{\rho\left( R_0^2 + \frac{4\eta R_o}{c\rho} \right)}
\frac{1}{{(\omega_0^2-\omega^2)}^2 + \omega^2\beta^2}\nonumber\\
~&~&\times \left((\omega_0^2-\omega^2)\cos(\frac{\omega}{\omega_0}t+\theta) 
+ \omega\beta\sin(\frac{\omega}{\omega_0}t+\theta)\right) \\
v &=& \dot x =-\frac{P_s\left( 1 + \frac{R_0^2\omega^2}{c^2} \right)^{\frac{1}{2}} }
{\rho\left( R_0^2 + \frac{4\eta R_o}{c\rho} \right)}\frac{\omega}{\omega_0}ja
\frac{1}{{(\omega_0^2-\omega^2)}^2 + \omega^2\beta^2}\nonumber\\
~&~&\times \left(-(\omega_0^2-\omega^2)\sin(\frac{\omega}{\omega_0}t+\theta) 
+ \omega\beta\cos(\frac{\omega}{\omega_0}t+\theta)\right) \nonumber\\ 
\end{eqnarray}
where $\theta$ denotes the phase.\\ 
Combining eqns.(11) and (12) we obtain 
\begin{equation}
\frac{\omega^2}{\omega_0^2}x^2 + v^2 = \frac{\omega^2}{\omega_0^2}
\frac{P_s^2}{R_0^4\rho^2{(1+\frac{4\eta}{c\rho R_0})}^2}\
\frac{1}{\left({(\omega_0^2-\omega^2)}^2 + \omega^2\beta^2\right)}
\end{equation}
Again using eqn.(6) to rewrite ~$x=R/R_0 -1$ ,and ~$v=\dot R/R_0$ in eqn.(13), we obtain after some 
algebra an equation for $\omega$: 
\begin{equation}
\omega^4 + \frac{K}{\rho^2R_0^4{\left(1+\frac{4\eta}{c\rho R_0}\right)}^2}\omega^2 
+ \frac{G^2}{\rho^2R_0^4{\left(1+\frac{4\eta}{c\rho R_0}\right)}^2} = 0 ~~~~~~~~~~~~~. 
\end{equation} 
where 
\begin{eqnarray}
G &=& 5(P_0-P_v)+\frac{8\sigma}{R_0}-\frac{Q^2}{8\pi\epsilon R_0^4} \nonumber\\
K &=& -\frac{2R_0^2}{c^2}(c^2\rho + P_0-P_v)G 
+ \frac{R_0^2}{c^2}{\left (\frac{4\eta c}{R_0}-(P_0-P_v)\right)}^2 \nonumber\\
~&~&-\frac{R_0^6P_s^2\rho \left(1+\frac{4\eta}{c\rho R_0}\right)}
{\left[ \rho R_0^4\left(1+\frac{4\eta}{c\rho R_0}\right)(R^2-2RR_0+R_0^2)+G{\dot R}^2 \right]}
\end{eqnarray}
This leads to the following expression for $\omega$ 
\begin{eqnarray}
\omega &=& \frac{1}{\rho R_0^2\left(1+\frac{4\eta}{c\rho R_0}\right)} \nonumber\\
&~&\times {\left[-K \pm {\left(K^2-4\rho^2R_0^4 {\left(1+\frac{4\eta}{c\rho R_0}\right)}^2 G^2\right)}^{1/2}
\right]}^{1/2}
\end{eqnarray}
After a careful look at each of the terms in this equation, we find that the dominant contribution 
of $Q$ to $d\omega /dQ$ ~ occurs as a cubic term :
\begin{equation}
\frac{d\omega}{dQ} \sim a_3 Q^3, 
\end{equation}
$a_3$ being a prefactor with appropriate dimensions. 
Integrating both sides of this equation between the limits corresponding to $Q=0$ ~and ~$Q$ gives
\begin{equation}
\omega-\omega_H \sim a_3 Q^4,
\end{equation} 
where $\omega_H$ is the frequency for the bubble with zero charge at which the $\dot{R}_{max} = c_1$, 
so that 
\begin{equation}
Q \sim a_3^\prime{(\omega-\omega_H)}^{1/4}, 
\end{equation} 
(the prefactor $a_3^\prime$ having appropriate dimensions), reproducing eqn.(\ref{eq3}) that was 
obtained from an analysis of the numerical results shown in the plots in Figure (\ref{fig5}). 
Hence it is very easy to predict the minimum charge $Q_{min}$ required on a bubble at a given applied 
pressure amplitude for attaining some particular value of the bubble's radial velocity, once $\omega_H$ 
is known.\\ 
\begin{figure}[!h]
\includegraphics[height=7cm,width=6.5cm,angle=0]{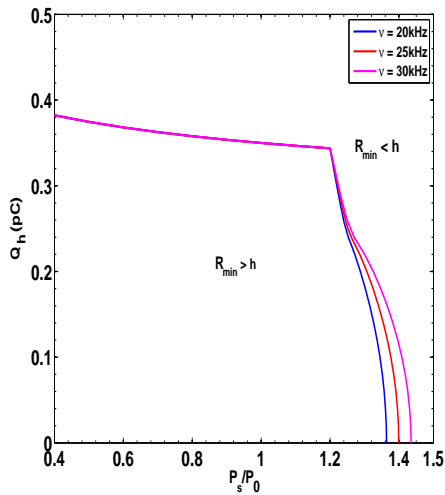}
\caption{$Q_{h}$ as function of $P_s$ for a bubble of $R_0=2\mu m$  for three different values 
of driving frequencies $\nu$. The value of $P_s$ at which $Q_h =0$, increases with increasing frequency. 
The physically reachable region is the area below the curve corresponding to $R_{min}>h$. 
The region above the curve corresponds to the (unphysical) regime where $R_{min}<h$. (color online).} 
\label{Qhplots}
\end{figure}
There is another interesting feature to be noted in this transition. The value of 
the frequency $\omega_H$ depends on the magnitude of the amplitude $P_s$ of 
the driving pressure wave. Indeed, for a given ambient bubble radius $R_0$, 
$\omega_H$ takes the simple linear form 
\begin{equation}
\omega_H =  b_1 + b_2 P_s,
\end{equation}  
where $b_1$ and $b_2$ vary with $R_0$. 
This can be seen clearly from the plot (Figure (\ref{fig5})c). 
A further functional dependence of $b_2$ on $R_0$, that is, of the slope $d\omega_H/dP_s$ 
on $R_0$, is also found (Figure (\ref{fig5})d), and is of the form 
\begin{equation}
\frac{d\omega_H}{dP_s} \sim R_0^{-0.9}.
\end{equation}
The maximal charge, $Q_{max}$, which a bubble can carry, is bounded by the fact 
that beyond a value $Q_h$ of the charge, bubble dimensions may reduce to below the value of the 
van der Waals hard core radius for the gas enclosed, which is physically untenable. 

Hence, the value of $Q_h$, the physically feasible maximal limit to the charge the bubble 
may carry, will be less than $Q_{max}$ for a particular $R_0$. 
Moreover, it depends as well on the amplitude of the forcing pressure $P_s$, with 
$Q_h$ decreasing with increasing $P_s$ and also with decreasing driving frequency. 
Figure (\ref{Qhplots}) show plots of $Q_h$ as a function of $P_s$ for three different 
driving frequencies, for $R_0=2\mu m$. Below a certain value of $P_s$, $Q_h$ becomes nearly 
independent of frequency as well as $P_s$. 

\section{Bifurcation diagrams} 
That the driving frequency influences the bubble dynamics is unquestionable. 
Techniques of dynamical systems theory have been used for long in the literature to understand 
bubble stability under variation of parameters (see for example~\cite{lauterborn, smereka, 
lauterborn2, holt, parlitz}.
Parlitz, et al.~\cite{parlitz} 
have, in their work, extensively investigated the frequency bifurcation diagrams for the bubble 
radius at various values of the driving amplitude pressure, $P_s$. \\

\begin{figure}[!h]
\includegraphics[height=8cm,width=8.5cm,angle=0.0]{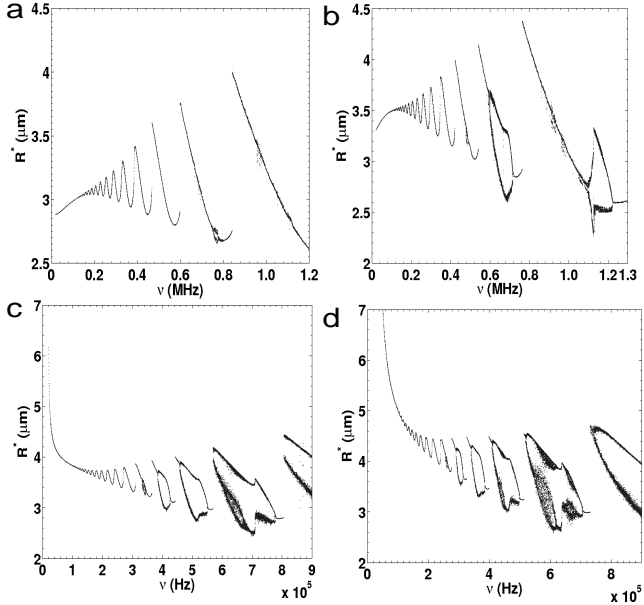}
\caption{Bifurcation diagram of a bubble oscillator with driving frequency as the control 
parameter, and for $R_0=2\mu m$ and (top): $P_s=1.2P_0$ (i.e., $P_s<P_{tr}$), and 
(bottom): $P_s=P_{tr}= 1.3P_0$. ~$Q=0$ in (a) and (c), and $Q=0.2pC$ in (b) and (d).}
\label{2micronbdPs12}
\end{figure}

\begin{figure}[]
\includegraphics[height=7.5cm,width=15cm,angle=270]{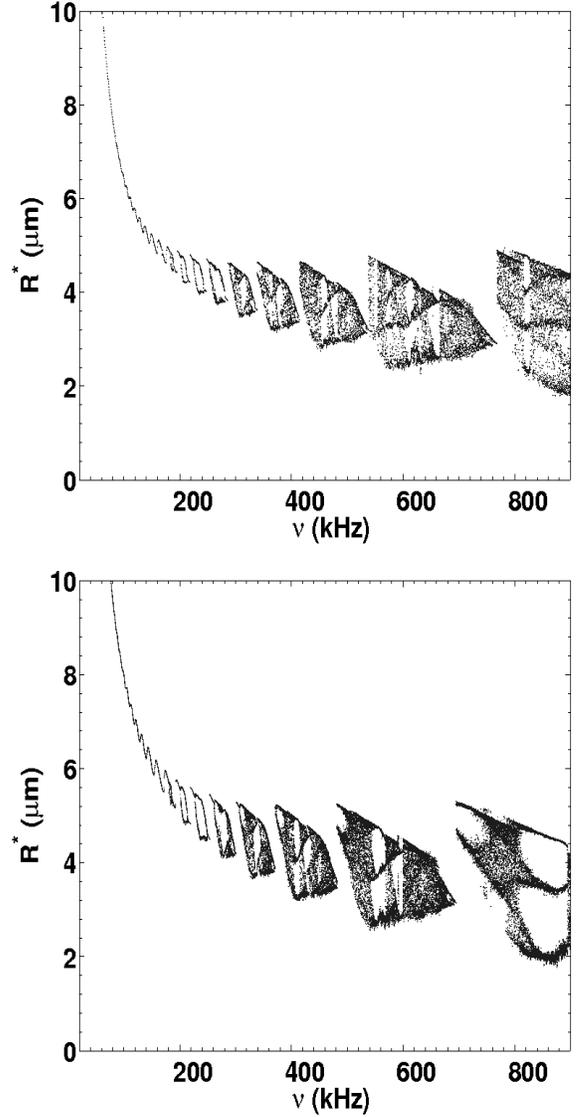}\\
\caption{Bifurcation diagram with respect to driving frequency of a bubble oscillator with $R_0=2\mu m$ 
and $P_s=1.4P_0$ for an uncharged bubble (top) and having charge $Q=0.2pC$ (bottom). 
(Note that $P_s>P_{tr}$.)} 
\label{2micronbdPs14}
\end{figure}

\begin{figure}[]
\includegraphics[height=8.2cm,width=8.2cm,angle=0]{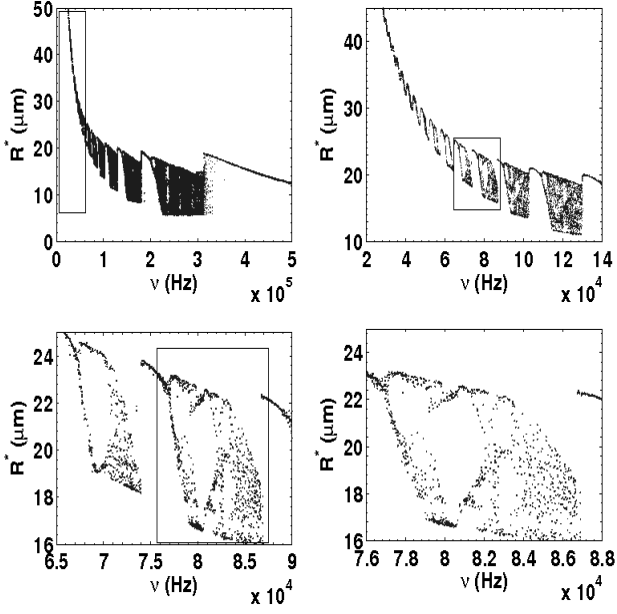}\\
\caption{Bifurcation diagram with respect to driving frequency of a bubble oscillator 
with $R_0=5\mu m$ and for $Q=0.8pC$. The driving pressure $P_s=1.4P_0$ (Here, $P_s>P_{tr}$). 
Successive images (b,c,d) show magnification of preceding insets showing period-doubling 
and chaos.}
\label{5micronbdPs14}
\end{figure}

\begin{figure}[]
\includegraphics[height=14cm,width=7.5cm,angle=0]{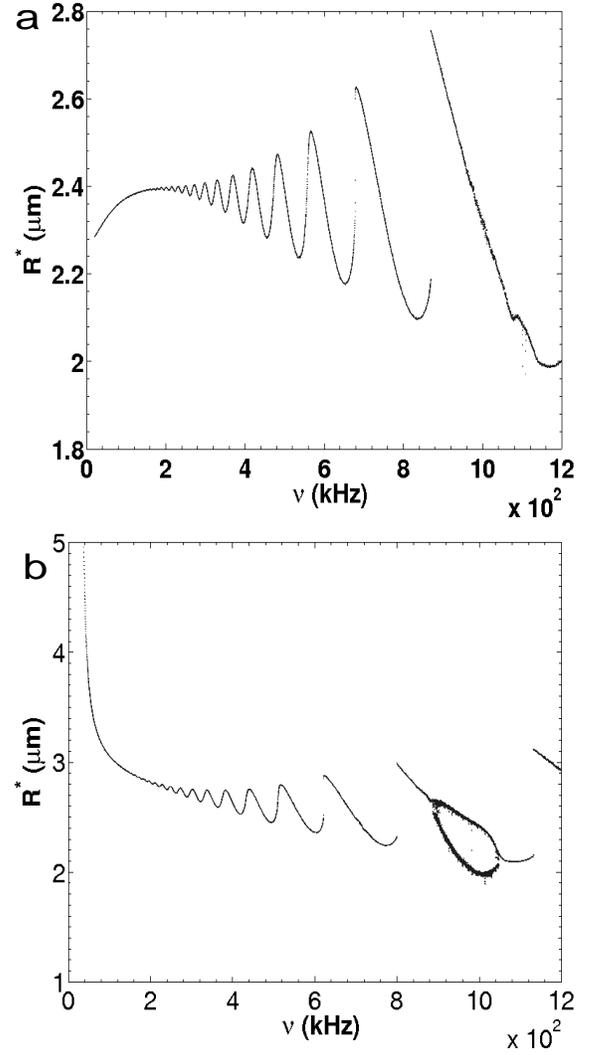}
\caption{Bifurcation diagram of a bubble oscillator with driving frequency as the control 
parameter for a smaller bubble. $R_0=1.45\mu m$ and $P_s=1.4P_0$.\\ 
(a) $Q=0$; ~(b) $Q=0.1pC$.} 
\label{bif145plot}
\end{figure}

In Figs.(\ref{2micronbdPs12}-\ref{bif145plot}) we have shown the bifurcation diagrams for the 
maximum radial amplitude $R_{max}$  of the given time series of the bubble, with the driving 
frequency as the control parameter for $R_0 = 1.45\mu m, 2\mu m$ and $5 \mu m$ for uncharged 
and charged bubbles. 

The bifurcation diagrams for various sets of parameters are constructed 
by sampling the time series after making sure the transients have decayed, 
for every time period $T=1/\nu$  of the external acoustic driving pressure 
elapsed. These sample points are precisely the points of intersection 
of the trajectories in phase plane with the Poincare cross section, and 
the orbit formed by the points represents the Poincare map. 
The bifurcation diagram is then constructed by plotting the sampled 
points calculated for a range of values of the control parameter (frequency 
or charge) and then plotting it with the control parameter on the 
horizontal axis and the sampled points on the vertical axis. 

In the response curves where period-doubling bifurcations occur, the branches always merge back to 
give period-1 oscillations. 

The presence of non-zero charge on the bubble advances period-doubling bifurcations with the driving 
frequency as the control parameter. This is demonstrated in the bifurcation diagrams in Figs.(8-9). 
For an uncharged bubble with ambient radius $R_0$ of 2microns at a driving pressure of 1.2$P_0$, 
period doubling is first seen at around 720kHz for the uncharged bubble, while the presence of 
0.2pC charge advances it to about 600 kHz (Fig.(\ref{2micronbdPs12}, a,b)). We observe that there 
are no chaotic regimes present at least till driving  frequencies of 1000kHz for low driving pressures 
such as this. 
 
Figs.(\ref{2micronbdPs12}-\ref{5micronbdPs14}) show that increasing the external pressure $P_s$ also 
has the effect of advancing the succession of period-doubling-period-halving bifurcations both for 
the charged as well as for the uncharged systems. For instance at 1.3$P_0$ 
(Figs.\ref{2micronbdPs12} c,d), the first period doubling bifurcation occurs at a forcing frequency 
of approximately 320kHz, followed by period halving bifurcation at 350kHz leaving period 1 oscillations, 
whereas on introduction of charge $Q=0.2$pC, the first period doubling bifurcation makes its appearance 
much earlier, at about 295 kHz, only to merge back to period 1 oscillations through a period halving 
bifurcation at 315 kHz. As one increases the driving frequency further, one observes the occurrence 
of a sequence of period-doubling - period-halving bifurcations.  

It should be noted that the chaotic regions make their appearance at the upper transient 
threshold pressure $P_{tr}$ (which for an uncharged bubble of $R_0=2\mu m$ is $1.3P_0$), and 
become more prominent for $P_s>P_{tr}$ (Fig.\ref{2micronbdPs14}). 
At large driving pressures, bands of chaotic regimes are present at high values of the forcing 
frequency, in agreement with observations of time series data. We show this in 
Figs.(\ref{2micronbdPs14}) for $P_s=1.4P_0$ : chaotic behaviour is seen to be present even at 
around 270-300kHz for charged and uncharged bubbles. \\
In Figs.(\ref{5micronbdPs14}) the sequence of period-2 and period-1 oscillations generated is shown 
for a slightly larger bubble, with $R_0=5\mu m$ and bearing charge $Q=0.8pC$, driven at pressure 
amplitude $P_s=1.4P_0$.\\

From these observations and other plots (not shown here) we deduce that the maximal radial 
amplitude of the bubble of a given equilibrium radius $R_0$  shows chaotic behaviour as a 
function of the driving frequency $\nu$, for $P_s \ge P_{tr}$ at large values of $\nu$. 
It was shown in ~\cite{ashok} that $P_{tr} = 1.12P_0$ for $R_0=5\mu m$  and $P_{tr} = 1.3P_0$ 
for $R_0=2\mu m$ for the uncharged bubble.\\   
 
For smaller frequencies, such as in the sonoluminescent regime, the presence of charges 
do not appear to introduce period-doublings in the system. However, the effect of charges in 
bringing about drastic changes in the bubble stability is more pronounced for smaller values 
of $R_0$. This is demonstrated in Figs.(11 a,b) for a bubble with $R_0= 1.45\mu m$ and driving 
pressure amplitude $P_s$ of $1.4P_0$. \\
The presence of 0.1pC charge on the bubble (Figs.(\ref{bif145plot}b)) induces a period-doubling 
bifurcation at a driving frequency of 880 kHz followed in quick succession by a period-halving 
bifurcation at 1040 kHz. These are absent for an uncharged bubble (Fig.(\ref{bif145plot}a)).\\
\begin{figure}[!h]
\includegraphics[height=7.5cm,width=7.5cm,angle=0]{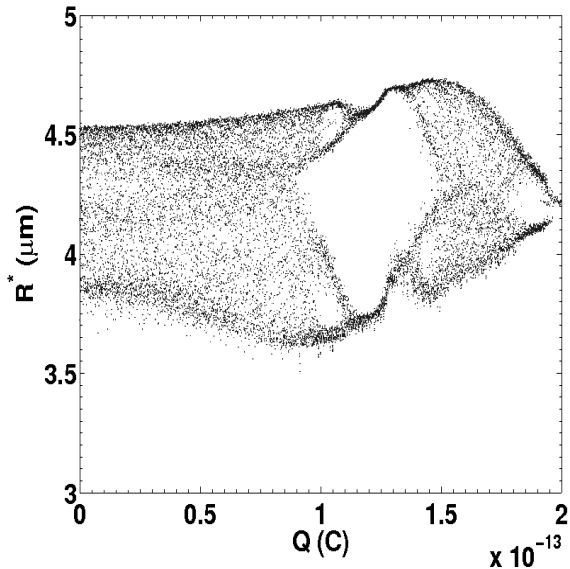}
\caption{Bifurcation diagram of a bubble oscillator with charge $Q$ as the control 
parameter, and with $R_0=2\mu m$ and $P_s=1.4P_0$ and at $\nu = 300kHz$. 
Non-chaotic windows are seen, most prominent one centred around $Q=0.125pC$.}
\label{chargebif2micronplot}
\end{figure}

In Fig.(\ref{chargebif2micronplot}) we obtain the bifurcation diagram of a bubble of $R_0=2\mu m$ 
driven by sonic pressure amplitude of $1.4P_0$ and frequency 300 kHz with charge as the control 
parameter. The choice of 300 kHz for the driving frequency has been made using the bifurcation 
diagram with frequency as the parameter (Fig.(9)) where the system is just beginning to get chaotic 
at this frequency. In Fig.(12) we see an interesting non-chaotic region centered around $Q= 0.125 pC$ 
with period-doubling and period-halving cascades. 
While constructing the bifurcation diagrams care has been taken to ensure that we work only within 
the range of charges permissible for a bubble of a given ambient radius. 
 
\section{The collapsing bubble: frequency \& charge dependence of temperature}
Investigating the maximum temperature as a function of the driving frequency, we obtain the interesting 
result that there exist two distinct domains of behavior of $T_{max}$ depending upon the amplitude 
of the driving pressure, $P_s$.\\
At lower pressures, i.e, for $P_s < P_{tr}$ (for example for $P_s = 1.1 P_0$ for $R_0 = 5 \mu$m), 
$T_{max}$ increases with driving frequency. However, as this value of $P_s$ falls in the vicinity 
of the transient threshold in the unstable regime (Fig.(4a)), we would expect 
$T_{max}$ to show large oscillations with frequency, as is also seen in Fig.(13a).  

At higher pressures, i.e, for $P_s > P_{tr}$ (for example for $P_s = 1.25 P_0$ for $R_0 = 5 \mu$m), 
the maximal temperature's frequency dependence is the opposite, with $T_{max}$ decreasing  uniformly 
with increasing frequency, showing oscillatory behavior (Fig.(13b)). $T_{max}$ shows a 
frequency-dependence of the form   
\begin{equation}
T_{max} = a_4 \times (\nu - a_5)^{-4/5},
\end{equation}
$a_4$ and $a_5$ being constants with appropriate dimensions. 
This is understood by recalling that the temperature is obtained from 
\begin{equation}
T = T(0) \left(\frac{R_0^3 - h^3}{R^3 - h^3}\right)^{2/3},
\label{TEqn}
\end{equation}
for $\Gamma = 5/3$. Making the approximation that $(h/R_0)^3 <1$ and also that $(h/R)^3 <1$ 
is sufficiently small at most values of $R$, we can approximate Eqn.(\ref{TEqn}) by 
\begin{equation}
T \approx T(0)R_0^2 / R^2.
\end{equation} 
Since at regimes at or near the Rayleigh collapse, $R(t) \sim \omega^{2/5}$, it immediately 
follows that 
\begin{equation}
T(t) \propto \omega^{-4/5}.
\end{equation}
Values of $T_{max}$ at lower pressures are less than that at higher pressures.  
\begin{figure}[!h]
\includegraphics[height=14cm,width=8.5cm,angle=0]{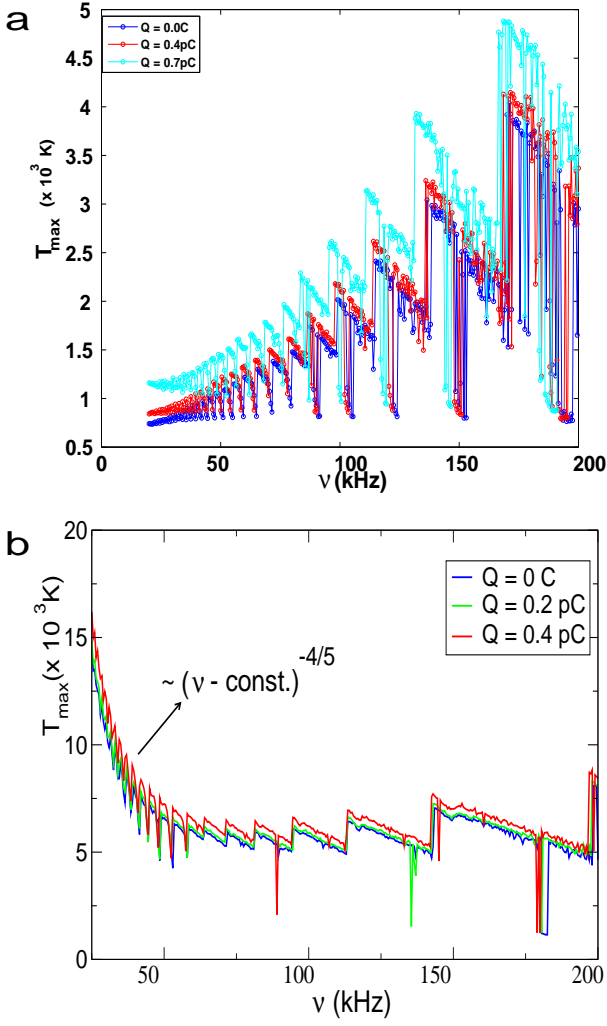}\\
\caption{Maximum temperature $T_{max}$ vs. $\nu$ for  different values of charge $Q$ at 
(a) $P_s = 1.1 P_0$ (lower / medium pressure,$P_s < P_{tr}$)  \& (b) $P_s = 1.25 P_0$ (high 
pressure, $P_s>P_{tr}$); the contrasting behavior above and below $P_{tr}$ can be seen. (color online).}
\label{fig13}
\end{figure}
\begin{figure}[!h]
\includegraphics[height=14cm,width=7.5cm,angle=0]{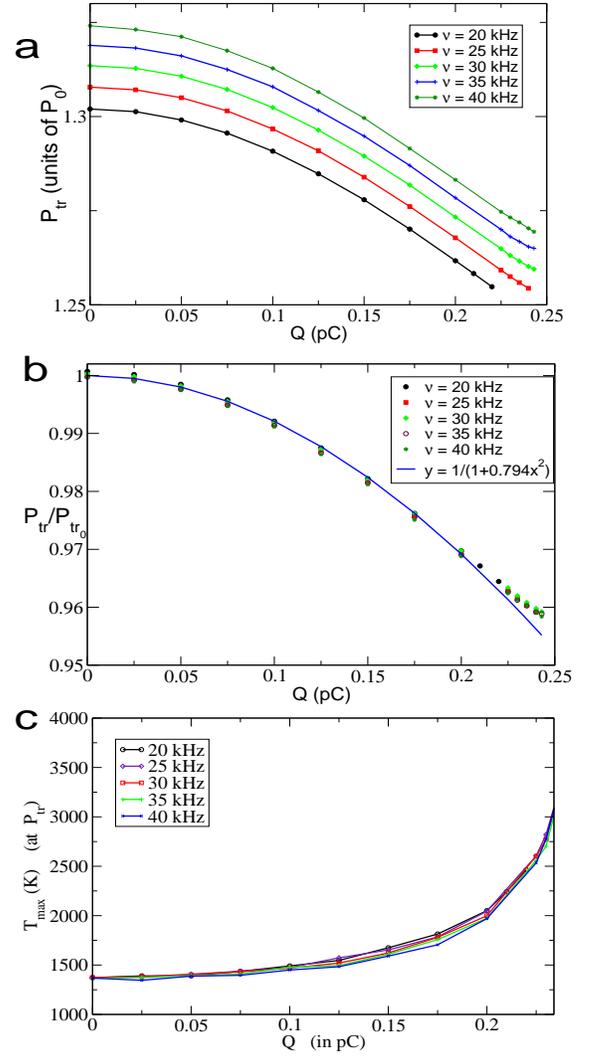}\\
\caption{(a) $P_{tr}$ decreases with increasing $Q$ over all driving frequencies. At a given charge, 
higher driving frequencies imply higher $P_{tr}$. (b) The $P_{tr}-Q$ curves when scaled by 
$P_{tr}$ at $Q=0$, ${P_{tr}}_0$, all fall on a master curve obeying eqn.(\ref{PtrMasterCurve}). 
(c) $T_{max}$ vs. $Q$ curves plotted at pressure value equalling the corresponding upper transient 
threshold value $P_{tr}$ for that particular value of $Q$, all fall nearly on the same curve. 
Plots are for $R_0 = 2 \mu$m. (color online).} 
\label{fig12new}
\end{figure}
Temperature $T_{max}$ rises steeply with pressure $P_s$ 
after some critical value of the pressure that equals the upper transient pressure threshold. 
Temperature calculations have also been reported by Wu and Roberts~\cite{wuRoberts} and 
Yasui~\cite{yasui} giving high magnitudes of the temperature. 
The presence of charge serves to reduce the pressure at which a maximum temperature $T_{max}$ 
is reached. As seen in Fig.(5b), increasing the driving frequency shifts the curves to the right, 
i.e., the same maximal bubble velocity $v_{max}$ that is obtained at some driving frequency $\omega_1$ 
for a pressure amplitude ${P_s}_1$ is reached for a higher frequency $\omega_2 > \omega_1$  only 
at a higher pressure ${P_s}_2 > {P_s}_1$.  Figs.(5b) and (14) where $v_{max}$ vs. $P_s$,  and 
$P_{tr}$ vs. $Q$ and $T_{max}$ vs. $Q$ plots respectively are shown for $R_0= 5\mu$m and $2\mu$m 
for charged and uncharged bubble at different driving frequencies, illustrate the effect of driving 
frequency and charge on bubble velocity, temperature and the location of the upper transient 
pressure threshold. 

The net effect of charge is to raise the possible $T_{max}$ for a bubble in comparison to 
the uncharged bubble. A comparative plot of $T_{max}$ reached at the upper transient threshold 
pressure $P_{tr}$ as a function of charge $Q$ is shown for a few values of the driving frequency 
$\nu$ in Fig.(14).\\
The temperature $T_{max}$ increases with charge $Q$ for all forcing frequencies.

The value of $P_{tr}$ (for a given bubble-charge, $Q$) increase with $\nu$. $P_{tr}$ at a given 
driving frequency decreases with increasing $Q$. The dependence of $P_{tr}$ on $Q$ over all 
frequencies can be captured by a normalized plot of $P_{tr}/{P_{tr}}_0$ against $Q$, where 
$ {P_{tr}}_0$ is the upper transient threshold pressure  value at zero charge, for a given frequency. 
This yields a master curve approximately obeying a relation of the form
\begin{equation}
P_{tr}= \frac{{P_{tr}}_0}{1 + 0.794 Q^2}, 
\label{PtrMasterCurve}
\end{equation}
as seen in Fig.(14b). The maximal value of temperature $T_{max}$ reached in a driven oscillating 
bubble, at the corresponding, respective $P_{tr}$ (which varies with $\nu$ and $Q$), over all 
values of charge $Q$, seems almost independent of the driving frequency $\nu$, as shown in Fig.(14c).
 
At higher pressures, beyond the upper transient threshold pressure, bubble velocities become 
larger and of the order of $c$. We argue that the maximal kinetic energy 
$\frac{1}{2}M{\dot{R}}^2 \approx \frac{1}{2}Mc^2$ would  approximately equal the dominant 
electrostatic contribution to the potential energy $Q^2/(4\pi\epsilon R_0)$, so that using 
$M = 4\pi R_0^3\rho/3$, $Q^2 \sim R_0^4$.\\ 
This argument is independent of the driving frequency at which the bubble is being forced. 
At a sufficiently high charge $Q_{max}$, a bubble of ambient radius $R_0$ will collapse to the 
same minimum radius $R_{min}$, independent of the driving frequency of the forcing pressure amplitude. 
However, this $Q_{max}$ value will typically be greater than $Q_H$, the upper bound imposed on the 
charge $Q$ by the physically realistic requirement that $R_{min}$ does not go below the van der Waals 
hard-core radius. Thus while this results in $Q_H$ being the greatest, physically realistic value of 
charge that a bubble can carry, we can still read off the value of a larger $Q = Q_{max}$ from 
$R_{min}$ vs $Q$ plots at high driving pressures, by identifying the $Q$ at which frequency 
independence of the curves sets in and all the curves for different driving frequencies all converge 
to the same $R_{min}$. More detailed discussions of $R_{min}$ dependencies are included in our earlier 
work~\cite{ashok}, and we do not show the plots here.\\
Hence a comparative estimate of this maximal charge a bubble can carry, $Q_{max}$, for two different 
values of the initial bubble radius $R_0$, say, ${R_0}_a$ and ${R_0}_b$, would be obtained from
\begin{equation}
\left(\frac{{R_0}_a}{{R_0}_b}\right)^2 \sim \left(\frac{{Q_{max}}_a}{{Q_{max}}_b}\right),  
\end{equation}
(for larger bubbles, of order $O(\mu$m) and above).
This is essentially a statement that the maximal surface charge density a bubble can carry is 
approximately same regardless of its initial ambient radius for micrometer and larger bubble radius, 
given that all other system parameters like the surface tension, pressure conditions, viscosity, etc. 
remain unchanged, while the influence of the effective surface tension (including the correction for 
charge present) is predominant in the submicron range. Indeed, for sub-micron and at very small bubble 
sizes where the surface tension and charge terms become just  comparable, we would have instead 
(comparing electrostatic energy with surface tension or elastic energy) 
$Q^2/R_0\sim kR_0^2 = m\omega_b^2 R_0^3$, with $k = m\omega_b^2$ being an effective spring constant 
and $\omega_b$ some natural frequency and $m$ the oscillator mass, so that on substituting for 
$m= (4/3)\pi \rho R_0^3$, we get 
\begin{equation}
\left(\frac{{R_0}_a}{{R_0}_c}\right)^3 \sim \left(\frac{{Q_{max}}_a}{{Q_{max}}_c}\right)^2 
\end{equation} 
for two different values of the initial bubble radius $R_0$, ${R_0}_a$ and ${R_0}_c$.\\
That such a non-rigorous approach cannot give any accurate numbers is obvious. Nonetheless, it is 
useful in giving us rough estimates of the maximal charge that the bubble can carry, in the absence 
of a constraint such as that imposed by the van der Waals hard-core radius.\\
A comparison of the numbers so obtained in this rough and ready way to that obtained from the numerical 
results is given below in Table 1 for three different values of ambient bubble radius $R_0$, at 
$P_s = 1.35 P_0$. 
\begin{table}[!h]
\caption{$Q_{max}$ ratio from Eqn.(27) (through $R_0$ ratio) compared with numerical values of 
$Q_{max}$ from  data.}
 \begin{tabular}{|l|l|l|l|l|l|} 
\hline
\multicolumn{6}{|c|}{Table 1}\\
\hline
${R_0}_a$  & ${R_0}_b$ & ${Q_{max}}_a$ & ${Q_{max}}_b$ & $\left(\frac{{R_0}_a}{{R_0}_b}\right)^2$ & $\left(\frac{{Q_{max}}_a}{{Q_{max}}_b}\right)$\\
\hline
2 $\mu$m & 5 $\mu$m & 0.23 pC & 1.24 pC & 0.16 & 0.18\\
10 $\mu$m & 2 $\mu$m & 4.7 pC & 0.23 pC & 25.0 & 20.43\\
10 $\mu$m & 5 $\mu$m & 4.7 pC & 1.24 pC & 4.0 & 3.79\\
\hline
\end{tabular}
\end{table}

\section{Conclusions} 
The presence of charge on a bubble suspended in a fluid influences the bubble's oscillations under 
ultrasonic forcing, and some of the aspects of the dynamics have been addressed in this work, taking 
the polytropic constant $\Gamma = 5/3$ which governs the equation of state for adiabatic heat transfer. 
A dimensionless constant $\zeta$ which we introduced in an earlier work~\cite{ashok} helps us to 
identify clearly the Blake threshold and the upper transient threshold $P_{tr}$ for acoustic cavitation. 
We use this to understand the influence of driving pressure $P_s$ and frequency $\nu$ of the applied 
ultrasonic field on the bubble oscillations. The presence of charge reduces the effective surface 
tension on the bubble walls so that its maximum radius $R_{max}$ attained during the expansion phase 
is larger than when it is uncharged; similarly the minimum bubble radius during collapse $R_{min}$ 
is much smaller in magnitude when the bubble is charged. The charged bubble undergoes a more 
violent collapse, achieving far higher temperatures in its interior in comparison with the 
uncharged one. We find that when $P_s < P_{tr}$, the maximum temperatures $T_{max}$ achieved in 
the bubble increase with increasing $\nu$ and charge. For $P_s > P_{tr}$, $T_{max}$ 
obeys a power law decrease with respect to $\nu$, with an exponent of -4/5. The power law behaviour 
is also obtained analytically through scaling arguments near the regime of Rayleigh collapse.

Bifurcation diagrams of the maximal radial amplitude of the bubble as a function of the driving 
frequency show the presence of chaotic regimes for $P_s \ge P_{tr}$ for any given ambient bubble radius 
at fairly large driving frequencies. 
The route to chaos is through period-doubling followed by period-halving bifurcations. The effect 
of charge is to always advance these bifurcations. At the lower end of the ultrasound spectral range, 
for instance in the sonoluminescent regime, the presence of charges do not appear to induce any 
period-doublings.\\     
Consistent with the fact that the presence of charge has a greater dominating effect over surface 
tension on bubbles of smaller equilibrium radii~\cite{ashok}, the bifurcation diagrams demonstrate that 
the effect of charges in drastically changing bubble stability is more pronounced for smaller bubbles.\\ 
We obtain also the bifurcation diagram of the maximal radial amplitude at any given $P_s$ as a 
function of the charge at large driving frequency. Here too, period-doublings and period-halvings 
are seen interspersed with large chaotic regimes. 

We obtain analytically an estimate of the minimum charge $Q_{min}$ required on a bubble at a given 
magnitude of applied pressure to attain a certain value $c_1$ of the bubble radial velocity. 
We find that this is related by a simple power law to the driving frequency of the acoustic wave. 
We show that above a critical frequency $\omega_H$, uncharged bubbles necessarily have to oscillate 
at velocities below $c_1$. The calculations are reproduced numerically also. 
Further, $\omega_H$ depends upon $P_s$. 

\section*{Acknowledgments} T.H. acknowledges support through a Rajiv Gandhi National Fellowship 
from the University Grants Commission, New Delhi. 

\end{document}